\documentclass{phc-proc4-auth}
\usepackage{graphicx}
\usepackage{amsmath}

\begin{document}
\begin{frontmatter}
\title{The charge and spin sectors of the $t$-$t'$ Hubbard model}
\author{Adolfo Avella\corauthref{avella}}
\ead{avella@sa.infn.it} \ead[url]{http://scs.sa.infn.it/avella}
\corauth[avella]{Dipartimento di Fisica "E.R. Caianiello"
\protect\\ Unit\`a di Ricerca INFM di Salerno \protect\\
Universit\`{a} degli Studi di
Salerno \protect\\ Via S. Allende, I-84081 Baronissi (SA), Italy \protect\\
Tel.  +39 089 965418 \protect\\ Fax:  +39 089 965275}
\author{Ferdinando Mancini}
\ead{mancini@sa.infn.it} \ead[url]{http://scs.sa.infn.it/mancini}
\address{Dipartimento di Fisica ``E.R. Caianiello'' - Unit\`a INFM di Salerno \protect\\
Universit\`a degli Studi di Salerno, I-84081 Baronissi (SA),
Italy}
\begin{abstract}
The charge and spin sectors, which are intimately coupled to the
fermionic one, of the $t$-$t'$ Hubbard model have been computed
self-consistently within the two-pole approximation. The relevant
unknown correlators appearing in the causal bosonic propagators
have been computed by enforcing the constraints dictated by the
hydrodynamics and the algebra of the composite operators coming
into play. The proposed scheme of approximation extends previous
calculations made for the fermionic sector of the $t$-$t'$ Hubbard
model and the bosonic sector of the Hubbard model, which showed to
be very effective to describe the overdoped region of cuprates
(the former) and the magnetic response of their parent compounds
(the latter).
\end{abstract}
\end{frontmatter}

Quite recently, we have shown how it is possible to capture, by
means of the $t$-$t'$ Hubbard model, the single-particle
properties of cuprate materials in their overdoped region within
an approximation scheme that has ingredients like: a two-pole
reduction of the fermionic retarded propagator, the use of
composite operators and the implementation of algebraic
constraints \cite{Avella:98c,Avella:99}. On the other hand, we
have also shown how it is possible to reproduce the magnetic
response of the parent compounds of the cuprates within a similar
approach applied to the relevant bosonic causal Green's functions
\cite{Avella:03,Avella:03a}. Both of these schemes have proved to
be very reliable through many positive comparisons to existing
numerical data \cite{Avella:98c,Avella:03}.

Along this line, we decided to study the possibility to extend the
latter approach to the $t$-$t'$ Hubbard model in order to provide
further elements of analysis to our study of the cuprate
properties and, in particular, of the magnetic properties in the
underdoped region. The 2D $t$-$t'$ Hubbard model is described by
the Hamiltonian
\begin{equation}
H=\sum_{\bf i,j}(t_{\bf ij}-\mu \delta_{\bf ij})c^\dagger ({\bf
i},t)c({\bf j},t)+U\sum_{\bf i} n_\uparrow (i) n_\downarrow (i)
\end{equation}
We use the standard notation: $c(i)$, $c^\dagger (i)$ are
annihilation and creation operators of electrons in the spinorial
notation; ${\bf i}$ stays for the lattice vector and $i = ({\bf
i},t)$; $\mu$ is the chemical potential; $t_{\bf ij}$ denotes the
transfer integral; $U$ is the screened Coulomb potential;
$n_\sigma (i)=c_\sigma ^\dagger (i)c_\sigma (i)$ is the charge
density of electrons at the site {\bf i} with spin $\sigma$. For a
cubic lattice the hopping matrix has the form $t_{\bf
ij}=-4t\alpha_{\bf ij}-4t'\beta_{\bf ij}$, where $\alpha_{ij}$ and
$\beta_{ij}$ are the first and second neighbor projection
operators, respectively,
\begin{align}
&\alpha_{\bf ij}={1 \over 2N}\sum_{\bf k} {e^{{\rm i}{\bf k}
(R_{\bf i}- R_{\bf j})}}[\cos (k_x)+\cos (k_y)]\\
&\beta_{\bf ij}={1 \over N}\sum_{\bf k} {e^{{\rm i}{\bf k}
(R_{\bf i}- R_{\bf j})}}\cos (k_x)\cos (k_y)
\end{align}
We choose as fermionic basis the following doublet
\begin{equation}\label{basis}
\psi (i)=\left(
\begin{matrix}
\xi (i)\\
\eta (i)
\end{matrix}
\right)
\end{equation}
where $\xi (i)=[1-n(i)]c(i)$ and $\eta(i)=n(i)c(i)$ are the
Hubbard operators, and $ n(i)=\sum_\sigma n_\sigma(i)$. In the
two-pole approximation \cite{Avella:98c} the retarded GF $G(i,j)=
\langle R[\psi (i)\psi^\dagger (j)] \rangle $ satisfies the
equation
\begin{equation}
[\omega -\varepsilon ({\bf k})]G(k,\omega )=I({\bf k})
\end{equation}
where $I({\bf k})=F.T. \langle \{ \psi ({\bf i},t),\psi ^\dagger
({\bf j},t)\} \rangle $ and $ \varepsilon ({\bf k})=F.T. \langle
\{ {\rm i}{{\partial \psi ({\bf i},t)} \over {\partial t}},\psi
^\dagger ({\bf j},t)\} \rangle I^{-1}({\bf k})$; the symbol $F.T.$
denotes the Fourier transform. In the paramagnetic phase the
energy matrix $\varepsilon ({\bf k})$ depends on the following set
of internal parameters: $\mu$, $\Delta = \langle \xi ^\alpha
(i)\xi ^\dagger (i) \rangle- \langle \eta ^\alpha (i)\eta ^\dagger
(i) \rangle$, $\Delta' = \langle \xi ^\beta (i)\xi ^\dagger (i)
\rangle- \langle \eta ^\beta (i)\eta ^\dagger (i) \rangle$, $p=
\langle n_\mu ^\alpha (i)n_\mu (i) \rangle /4- \langle [c_\uparrow
(i)c_\downarrow (i)]^\alpha c_\downarrow ^\dagger (i)c_\uparrow
^\dagger (i) \rangle$, $p'= \langle n_\mu ^\beta (i)n_\mu (i)
\rangle /4- \langle [c_\uparrow (i)c_\downarrow (i)]^\beta
c_\downarrow ^\dagger (i)c_\uparrow ^\dagger (i) \rangle$, which
must be self-consistently determined. Given an operator
$\zeta(i)$, we are using the notation $\zeta ^\gamma (i)=\sum_{\bf
j} \gamma _{\bf ij} \zeta ({\bf j},t)$ and $\zeta ^{\gamma\lambda}
(i)=\sum_{\bf jl} \gamma _{\bf ij}\lambda_{\bf jl} \zeta ({\bf
l},t)$ with $\gamma,\lambda=\alpha,\beta$. The operator $n_\mu
(i)=c^\dagger (i) \sigma _\mu c(i)$ [$\sigma _\mu =({\bf
1},\mathbf{\sigma} )$, $\sigma$ are the Pauli matrices] is the
charge ($\mu = 0$) and spin ($\mu = 1,2,3$) density operator. The
local algebra satisfied by the fermionic field (\ref{basis})
imposes the constraint $\langle \xi (i)\eta ^\dagger (i) \rangle
=0$: this equation allows us to solve self-consistently the
fermionic sector \cite{Avella:98c}.

We consider then the composite bosonic field \cite{Avella:03}
\begin{equation}
N^{(\mu )}(i)=\left(
\begin{matrix}
n_\mu (i)\\
\rho_\mu (i)
\end{matrix}
\right)
\end{equation}
where $\rho_\mu(i)=\rho'_\mu(i)+\tau\rho''_\mu(i)$,
$\rho'_\mu(i)=c^\dagger (i)\sigma _\mu c^\alpha (i)-c^{\alpha
\dagger} (i)\sigma _\mu c(i)$ and $\rho''_\mu(i)=c^\dagger
(i)\sigma _\mu c^\beta (i)-c^{\beta \dagger} (i)\sigma _\mu c(i)$
with $\tau=t'/t$. The equation of motion of $n_{\mu}(i)$ reads as
$\mathrm{i}\frac{\partial}{\partial t}n_{\mu}(i)
=-4t\rho_{\mu}(i)$, whereas that of $\rho_{\mu}(i)$ reads as
$\mathrm{i}\frac{\partial}{\partial t}\rho_{\mu}(i) =-4tl_{\mu}(i)
+U k_{\mu}(i)$. According to the fact that $\tau$ is usually
chosen of the order $10^{-1}$, we have decided to neglect terms of
the order $\tau^2$ and higher, as they will practically give no
relevant contributions to the dynamics. In this case, we have
$l_{\mu}(i)  =l_{\mu}^{\prime}(i) +2 \tau
l_{\mu}^{\prime\prime}(i)$ and $\kappa_{\mu}(i)
=\kappa_{\mu}^{\prime}(i) + \tau \kappa_{\mu}^{\prime\prime}(i)$
with
\begin{align}
l_{\mu}^{\prime}(i)&  =c^{\dagger}(i) \sigma_{\mu}
c^{\alpha^{2}}(i)  +c^{\dagger\alpha^{2} }(i)   \sigma_{\mu}  c(i)
-2c^{\dagger
\alpha}(i)   \sigma_{\mu}  c^{\alpha}(i) \\
l_{\mu}^{\prime\prime}(i) & =c^{\dagger}(i)  \sigma_{\mu}
c^{\alpha\beta}(i) +c^{\dagger\alpha \beta}(i) \sigma_{\mu}  c(i)
-c^{\dagger\alpha}(i) \sigma_{\mu}  c^{\beta}(i)
\notag\\
&-c^{\dagger\beta}(i) \sigma_{\mu}  c^{\alpha
}(i) \\
\kappa_{\mu}^{\prime}(i)&  =c^{\dagger}(i) \sigma_{\mu}
\eta^{\alpha}(i) -\eta^{\dagger}(i) \sigma_{\mu}  c^{\alpha}(i)
+\eta ^{\dagger\alpha}(i) \sigma_{\mu}  c(i) \notag\\ &
-c^{\dagger\alpha}(i) \sigma_{\mu} \eta(i) \\
\kappa_{\mu}^{\prime\prime}(i)&  =c^{\dagger}(i) \sigma_{\mu}
\eta^{\beta}(i) -\eta^{\dagger}(i) \sigma_{\mu}  c^{\beta}(i)
+\eta ^{\dagger\beta}(i) \sigma_{\mu}  c(i)\notag\\ &
-c^{\dagger\beta}(i) \sigma_{\mu} \eta(i)
\end{align}

In the two-pole approximation the causal GF $G^{(\mu )}(i,j)=
\langle T[N^{(\mu )}(i)N^{(\mu )\dagger} (j)] \rangle $ satisfies
the equation
\begin{equation}
[\omega -\varepsilon^{(\mu )} ({\bf k})]G^{(\mu )}({\bf k},\omega
)=I^{(\mu )}({\bf k})
\end{equation}
where $I^{(\mu )}({\bf k})=F.T. \langle [N^{(\mu )}({\bf
i},t),N^{(\mu )\dagger}({\bf j},t)] \rangle $, $m^{(\mu )}({\bf
k})=F.T. \langle [ {\rm i}{{\partial N^{(\mu )}({\bf i},t)} \over
{\partial t}},N^{(\mu )\dagger} ({\bf j},t)] \rangle$ and $
\varepsilon^{(\mu )} ({\bf k})= m^{(\mu )}({\bf k}) [I^{(\mu
)}({\bf k})]^{-1}$. As it can be easily verified, in the
paramagnetic phase the normalization matrix $I^{(\mu )}$ does not
depend on the index $\mu$; charge and spin operators have the same
weight. The two matrices $I^{(\mu )}$ and $m^{(\mu )}$ are
off-diagonal and diagonal, respectively, and have the following
entries:
\begin{align}
& I^{(\mu )}_{12}(\mathbf{k})    =4\left[  1-\alpha\left(
\mathbf{k} \right)  \right]  C_{cc}^{\alpha}+4 \left[
1-\beta(\mathbf{k})  \right]  \tau C_{cc}^{\beta}\\
& m^{(\mu )}_{11}(\mathbf{k})     =-4 t I_{12b}(\mathbf{k})  \\
& m^{(\mu )}_{22}(\mathbf{k})    =-4 t I_{l_{\mu}\rho_{\mu}}(\mathbf{k})  +U I_{\kappa_{\mu}\rho_{\mu}}(\mathbf{k})  \\
& C_{cc}^\gamma = \left\langle c^\gamma ({\bf i})  c^\dagger ({\bf
i})\right\rangle\\
& I_{l_{\mu}\rho_{\mu}}(\mathbf{k}) = F.T. \langle [l_{\mu}({\bf
i},t),\rho_{\mu}^{\dagger}({\bf j},t)] \rangle\\
& I_{\kappa_{\mu }\rho_{\mu}}(\mathbf{k}) = F.T. \langle
[\kappa_{\mu }({\bf i},t),\rho_{\mu}^{\dagger}({\bf j},t)] \rangle
\end{align}
The energy matrix $\varepsilon^{(\mu )}(\mathbf{k})$ has
off-diagonal form with entries: $\varepsilon^{(\mu
)}_{12}(\mathbf{k})=-4 t$ and $\varepsilon^{(\mu
)}_{21}(\mathbf{k})=m^{(\mu )}_{22}(\mathbf{k})/I^{(\mu
)}_{12}(\mathbf{k})$. Quite lengthy calculations shows that the
energy spectra [$E_n^{(\mu )}(\mathbf{k})=(-)^n\sqrt{-4t m^{(\mu
)}_{22}(\mathbf{k})/I^{(\mu )}_{12}(\mathbf{k})} \quad n=1,2$] and
the spectral weights (which can be expressed in terms of the
eigenenergies and eigenvectors of the energy matrix) depend on the
following parameters: fermionic correlators $C_{nm}({\bf
i-j})=\left\langle\psi_n ({\bf i})\psi_m^\dagger ({\bf
j})\right\rangle$ with $|{\bf i-j}|$ up to $4$ lattice {\em hops}
along the two main axis of the lattice and unknown bosonic
correlators $a_\mu$, $b_\mu$, $c_\mu$ and $d_\mu$ (whose explicit
expression can be found in Ref.~\cite{Avella:03}) and \small
\begin{align}
a_{\mu}^{\prime} &  =\left\langle c^{\dagger\beta}(i)
c^{\alpha}(i)   n(i)  \right\rangle -\frac{1} {3}\left(
4\delta_{\mu0}-1\right)  \left\langle c^{\dagger\beta}(i)
\sigma_{q} c^{\alpha}(i)  n_{q}(i)
\right\rangle \nonumber\\
&  +4\left(  2\delta_{\mu0}-1\right)  \left\langle c_{\uparrow}(i)
c_{\downarrow}(i)  c_{\downarrow}^{\dagger\alpha }(i)
c_{\uparrow}^{\dagger\beta}(i)
\right\rangle \\
c_{\mu}^{\prime} &  =\frac{1}{4}\left\langle c^{\dagger}(i)
c\left(  i_{1}\right)   n\left(  i_{5}\right) \right\rangle
-\frac{1}{12}\left(  4\delta_{\mu0}-1\right) \left\langle
c^{\dagger}(i)   \sigma_{q} c\left( i_{1}\right)  n_{q}\left(
i_{5}\right)
\right\rangle \nonumber\\
&  +\left(  2\delta_{\mu0}-1\right)  \left\langle
c_{\uparrow}\left( i_{5}\right)   c_{\downarrow}\left(
i_{5}\right)   c_{\downarrow}^{\dagger }\left(  i_{1}\right)
c_{\uparrow}^{\dagger}(i)  \right\rangle\nonumber\\
&   +\frac{1}{4}\left\langle c^{\dagger}(i) c\left(  i_{5}\right)
n\left( i_{1}\right) \right\rangle -\frac{1}{12}\left(
4\delta_{\mu0}-1\right) \left\langle c^{\dagger}(i)  \sigma_{q}
c\left(  i_{5}\right) n_{q}\left( i_{1}\right)
\right\rangle \nonumber\\
&  +\left(  2\delta_{\mu0}-1\right)  \left\langle
c_{\uparrow}\left( i_{1}\right)   c_{\downarrow}\left(
i_{1}\right)   c_{\downarrow}^{\dagger }\left(  i_{5}\right)
c_{\uparrow}^{\dagger}(i)  \right\rangle
\\ d_{\mu}^{\prime} &  =\frac{1}{4}\left\langle c^{\dagger}(i)
c\left(  i_{1}\right)   n\left(  i_{13}\right) \right\rangle
-\frac{1}{12}\left(  4\delta_{\mu0}-1\right) \left\langle
c^{\dagger}(i)   \sigma_{q} c\left( i_{1}\right)  n_{q}\left(
i_{13}\right)
\right\rangle \nonumber\\
&  +\left(  2\delta_{\mu0}-1\right)  \left\langle
c_{\uparrow}\left( i_{13}\right)   c_{\downarrow}\left(
i_{13}\right)   c_{\downarrow }^{\dagger}\left(  i_{1}\right)
c_{\uparrow}^{\dagger}(i)
\right\rangle \nonumber\\
&  +\frac{1}{4}\left\langle c^{\dagger}(i) c\left(  i_{13}\right)
n\left(  i_{1}\right) \right\rangle -\frac{1}{12}\left(
4\delta_{\mu0}-1\right) \left\langle c^{\dagger}(i)  \sigma_{q}
c\left( i_{13}\right)   n_{q}\left( i_{1}\right)
\right\rangle \nonumber\\
&  +\left(  2\delta_{\mu0}-1\right)  \left\langle
c_{\uparrow}\left( i_{1}\right)   c_{\downarrow}\left(
i_{1}\right)   c_{\downarrow}^{\dagger }\left(  i_{13}\right)
 c_{\uparrow}^{\dagger}(i)
\right\rangle
\end{align}
\normalsize

In order to compute the unknown bosonic correlators we can resort
to the hydrodynamic constraints that require that the bosonic
spectra should be superlinear in momentum for long wavelength and
that the susceptibility should be single valued at ${\bf k=0}$.
Moreover, we can impose the local algebra constraint $\langle
n_\mu (i)n_\mu (i) \rangle = \langle n \rangle +2(2\delta _{\mu
0}-1)D$, where $D=\langle n \rangle/2-C_{22}({\bf 0})$ is the
double occupancy. According to this, we get four equations which
allow to compute the parameters $a_\mu$, $b_\mu$, $c_\mu$ and
$d_\mu$ in terms of $a'_\mu$, $c'_\mu$ and $d'_\mu$. These latter
parameters, with respect to the {\em no-primed} ones, are made up
of correlators containing operators centered at more distant
sites. According to this, we expect that their values and
relevance should be lower and in order to determine them we
suggest the following decouplings
\begin{align}
a'_\mu  &  \approx 8\left(  2\delta_{\mu0}-1\right) C_{cc}^{\alpha
}C_{cc}^{\beta}-nC_{cc}^{\alpha}-nC_{cc}^{\kappa}\\
c'_\mu  &  \approx 2\left(  2\delta_{\mu0}-1\right) C_{cc}^{\alpha
}\left(  C_{cc}^{\alpha}+C_{cc}^{\beta}\right)
-\frac{n}{2}C_{cc}^{\alpha
}-\frac{n}{2}C_{cc}^{\beta}\\
d'_\mu  &  \approx 2\left(  2\delta_{\mu0}-1\right) C_{cc}^{\beta
}\left(  C_{cc}^{\alpha}+C_{cc}^{\kappa}\right)
-\frac{n}{2}C_{cc}^{\alpha }-\frac{n}{2}C_{cc}^{\kappa}
\end{align}
where $\kappa$ is the projector on the fourth nearest neighbors.

In conclusion, we have reported a self-consistent scheme of
calculations for the (spin and charge) bosonic sector of the
$t$-$t'$ Hubbard model. It is worth noticing that, within this
scheme, the hydrodynamic constrains and the local algebra is
preserved assuring that the known limits are conserved. Results of
the presented scheme, easily attainable by solving numerically the
self-consistent equations, will be presented elsewhere.


\end{document}